\documentstyle[11pt,newpasp,twoside,psfig]{article}
\begin{document}

\title{An Infrared View of Local Universe AGN}
\author{Almudena Alonso-Herrero}
\affil{Steward Observatory, University of Arizona, Tucson, AZ85721, USA}

\setcounter{page}{111}
\index{Alonso-Herrero, A.}

\begin{abstract}
We summarize recent results on the infrared (IR) nuclear 
properties of Seyfert galaxies. 
100\% of all Seyfert $1-1.9$s  and 
$\simeq 50\%$ of all Seyfert 2s in the CfA and RSA 
samples observed with {\it HST}/NICMOS 
show nuclear point sources at $1.6\,\mu$m. 
We find that the unresolved emission is variable in 9 out of
the 14 Seyferts with duplicate observations, indicating a non-stellar  
origin. We have 
also put together non-stellar $0.4-16\,\mu$m
spectral energy distributions (SEDs)
for the entire CfA sample of Seyfert galaxies. The shape of the SEDs for 
a complete sample of AGN can be used to constrain torus models.
We report a good correlation between  the IR and hard X-ray 
luminosities of  local universe AGN suggesting that 
their mid-IR emission  
 is approximately an isotropic property, 
and thus can be used as an indicator of the AGN luminosity. 
Finally we show that 
the hard X-ray and mid-IR luminosities appear to be related to the 
black hole mass in AGN.

\end{abstract}

\section{Introduction}
The advent of new infrared (IR) facilities 
(both space and ground-based) with superb spatial resolution and sensitivity
is allowing us to make significant progress toward the understanding 
of the IR nuclear properties of local universe AGN. 
The IR emission in AGNs (Seyfert  galaxies and 
radio quiet quasars) is often interpreted 
as produced by hot dust in a torus/disk configuration  
(e.g., Barvainis 1987; Sanders et al. 1989; Pier \& Krolik 1993).  
However, until recently, the nuclear non-stellar 
IR continuum of Seyfert 2 galaxies has
eluded us because of the dominance of the stellar emission  
shortward of $3\,\mu$m (e.g., Alonso-Herrero et al. 1996).
In this paper we
 summarize recent results obtained by our group on the IR 
nuclear properties of Seyfert galaxies in the local universe.

\section{Unresolved Continuum Sources at $1.6\,\mu$m}

We have analyzed {\it HST}/NICMOS $1.6\,\mu$m observations
of 112 Seyfert galaxies and  found that 
$\simeq 50$\% of the Seyfert 2s in the Revised Shapely-Ames (RSA) Catalog
and  CfA redshift sample contain unresolved nuclear continuum sources 
at $1.6\,\mu$m. Most Seyfert $1-1.9$s 
display unresolved nuclear continuum sources at this wavelength. 
These unresolved sources have
$1.6\,\mu$m fluxes of order 1\,mJy, near-IR luminosities of order 
$10^{41}\,$erg\,s$^{-1}$, and absolute magnitudes 
$M_H \simeq -16$.  
Non-Seyfert galaxies from the RSA Catalog 
display significantly fewer
($\simeq 20$\%) nuclear unresolved 
sources with lower luminosities, which could 
be due to compact star clusters. We have also found that the luminosities of 
the unresolved Seyfert $1.0-1.9$ sources at $1.6\,\mu$m are correlated
with the [O\,{\sc iii}]$\lambda$5007 and hard ($2-10\,$keV) 
X-ray luminosities, implying that 
these sources are non-stellar.  

We find no strong correlation between the unresolved $1.6\,\mu$m fluxes
and hard X-ray or [O\,{\sc iii}]$\lambda$5007 fluxes for the 
pure Seyfert 2s. These galaxies also tend to 
have lower $1.6\,\mu$m luminosities compared to the Seyfert 
$1.0-1.9$ galaxies of similar [O\,{\sc iii}] luminosity. This is in 
agreement with the findings of Alonso-Herrero
et al. (1997) using $L$ ($3.5\,\mu$m) nuclear fluxes. 
Either large extinctions 
($A_V \simeq 20-40\,$mag) are present toward their continuum-emitting 
regions or some fraction of the unresolved sources at $1.6\,\mu$m 
in Seyfert 2s 
are compact star clusters (see Quillen et al. 2001 for more details). 

Further evidence for the non-stellar
origin of the unresolved nuclear emission at  $1.6\,\mu$m comes from 
variability shown by 9 of the 14 nuclei studied in Quillen et al. (2000).  
The variability is at the level 
of $10\%-40\%$ on timescales of $\simeq 1-14$ 
months, depending upon the
galaxy. A control sample of Seyfert galaxies lacking unresolved 
sources and galaxies lacking Seyfert nuclei show less than 3\% instrumental
variation in equivalent aperture measurements. 

\begin{figure}[h]
\plotfiddle{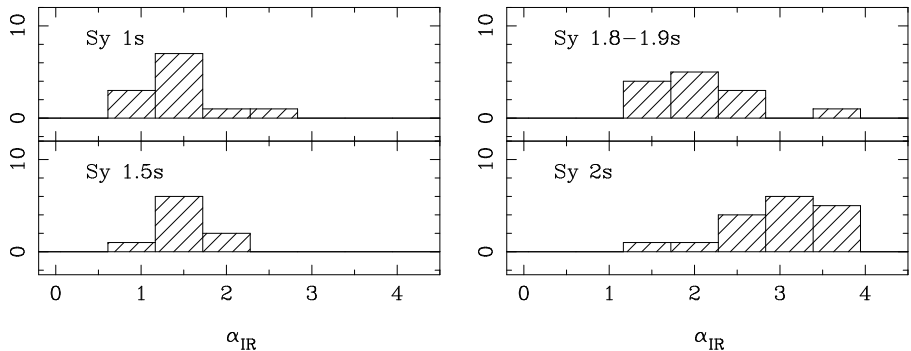}{425pt}{0}{70}{70}{-230}{260}
\vspace{-10.5cm}
\caption{Distributions of the IR ($1-16\,\micron$) 
spectral indices ($f_\nu \propto \nu^{-\alpha_{\rm IR}}$) 
for 51 Seyferts in the  
CfA sample. The sample has been 
broken up in different Seyfert types as derived 
from optical spectroscopy.}
\end{figure}

\section{SEDs of the CfA Seyfert galaxies}

We have put together nuclear SEDs
in the  $0.4-16\,\micron$ spectral range 
for the entire CfA
sample of Seyfert galaxies,  including those LINERs reclassified
as Seyferts by later spectroscopic studies (see Alonso-Herrero et al. 2002b).
We find that the spectral indices ($f_\nu \propto \nu^{-\alpha_{\rm IR}}$) 
of the unresolved AGN emission  
in the region $1-16\,\micron$ vary from $\alpha_{\rm IR} \sim$ 0.9 to 3.8.   
The shapes of the spectra are correlated with
the Seyfert type in the sense that steeper nuclear SEDs ($\nu f_\nu$ increasing
with increasing wavelength) tend
to be found in Seyfert 2s and flatter SEDs 
($\nu f_\nu \simeq$ constant) are in the Seyfert $1-1.5$s (see Fig.~1).
The majority
of galaxies optically classified as Seyferts 1.8s and 
1.9s display values of $\alpha_{\rm IR}$ either 
as in  type 1 objects (mean for type 1s $\alpha_{\rm IR} = 1.5\pm0.4$), or
intermediate between Seyfert 1s and Seyfert 2s (Fig.~1). 
The SED of intermediate Seyfert $1.8-1.9$s may be consistent with 
a pure Seyfert 1 SED viewed through moderate amounts 
($A_V < 5\,$mag) of foreground galaxy extinction. 

Torus models usually adopt high equatorial opacities  to 
reproduce the IR properties of Seyfert 1s and 2s
(e.g., Pier \&
Krolik 1993; Efstathiou \& Rowan-Robinson 1995), 
resulting in a dichotomy of SEDs ---flat for type 1s, and 
steep for type 2s. Such dichotomy is not observed in 
the CfA Seyferts, and in particular there is a wide range of  
observed $\alpha_{\rm IR}$ in 
type 2s (Fig.~1). The  lack of steep SEDs, and large numbers
of objects with intermediate spectral indices cannot be reconciled 
with predictions
from existing optically thick torus models. Possible 
modifications 
to existing torus models include low optical depth tori, 
clumpy dusty tori and 
high optical tori with an extended optically thin component.

\begin{figure}[h]
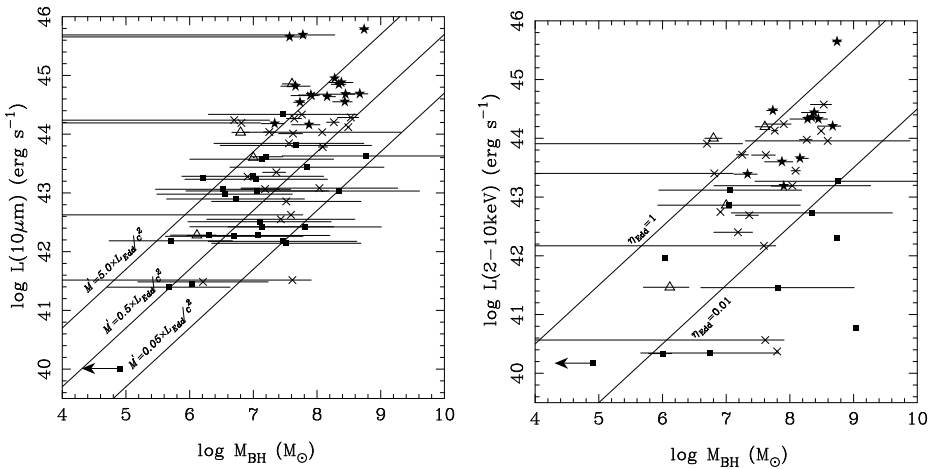

\plotfiddle{aalonso_fig2a.ps}{425pt}{-90}{40}{40}{-195}{480}
\plotfiddle{aalonso_fig2b.ps}{425pt}{-90}{40}{40}{-15}{915}
\vspace{-24cm}
\caption{{\it Left panel}: Relation between the $10\,\mu$m nuclear 
luminosity and the BH mass. Filled stars are PG quasars, crosses 
are Seyfert 1s, open 
triangles are narrow-line Seyfert 1s, and filled squares are Seyfert 2s. 
{\it Right panel}: Same as left panel, but for the hard X-ray 
($2-10\,$keV) luminosity and BH mass. }
\end{figure}

\section{Mid-IR emission of AGN, an indicator of the AGN 
luminosity and black hole mass?}

The hard X-ray ($2-10\,$keV) emission of Seyfert galaxies is known to 
be a good indicator of the intrinsic luminosity of the AGN for those cases
where it is transmitted through the torus, that is, in Compton thin galaxies. 
Other proposed isotropic (that is, not dependent on the viewing angle
to the AGN)
indicators of the AGN emission  in Seyfert 1s and Seyfert
2s include the [O\,{\sc iii}]$\lambda 5007$ luminosity 
(Mulchaey et al. 1994; Heckman 1995) or the nonthermal 
1.45 GHz radio continuum (Heckman 1995). 

We find a good correlation between ground-based 
$4.8\,\mu$m and {\it ISO} $9.7\,\mu$m and hard X-ray fluxes and 
luminosities for both Seyfert 1s and Compton thin
($N_H \le 10^{\rm 24}\,{\rm cm}^{-2}$) Seyfert 2s (see Alonso-Herrero
et al. 2001, 
also Krabbe et al. 2001). This indicates that the mid-IR	 
emission in Seyfert galaxies can be used as a measure of the AGN luminosity.
The improved 
correlations at 4.8 and $9.7\,\mu$m with respect to those at 
shorter wavelengths (Alonso-Herrero et al. 1997, and 
Section~2) are explained in
terms of the reduced extinction. Some Compton thick sources 
(e.g., NGC~1068 and Mrk~533) are bright IR sources suggesting that the
      component responsible for the bulk of the IR emission 
in Seyfert galaxies is more visible from all viewing angles than 
that responsible for
      the hard X-ray emission. 

We have also compiled a heterogeneous sample of local AGN that includes 
Seyfert 1 galaxies, Seyfert 2 galaxies, and
PG quasars to investigate for the first time the 
relation between black hole (BH) mass  and mid-IR nuclear 
emission (Fig.~2, left panel). We find
a clear relation between the BH mass and the $10\,\mu$m nuclear luminosity for 
these local AGNs. There are no significant differences between type 1 and
type 2 objects, implying that the reprocessing of the $10\,\mu$m 
nuclear emission is not severely affected by geometric and 
optical depth effects.
We also confirm that the BH mass is related to the $2-10\,$keV X-ray 
luminosity, but only for Compton-thin galaxies (Fig.~2, right panel). 
Our results show that rest-frame $10\,\mu$m and hard X-ray
luminosities (especially the former, which is applicable to 
all AGN types) may be used to derive
BH masses at high
redshift and to probe their cosmological evolution 
(see Alonso-Herrero et al. 2002a).

\acknowledgements
We are very grateful to A. C. Quillen, 
G. H. Rieke, M. J. Rieke, M. J. Ward, 
C. Simpson, V. D. Ivanov, A. Efstathiou, 
R. Jayawardhana, T. Hosokawa, S. Shaked, C. McDonald 
and A. Lee for their contributions to the results presented in 
this paper. 

AAH acknowledges support from NASA projects NAG-53359 and
NAG-53042 and from JPL Contract No.~961633.

\end{document}